\documentclass[aps,pre, groupedaddress,showpacs,twocolumn,amsmath,amssymb,floatfix]{revtex4-1}

\usepackage[dvipdfmx]{graphicx}
\newcommand{\VEC}[1]{\overrightarrow{#1}}
\newcommand{\pdd}[2]{\displaystyle \frac{\partial #1}{\partial #2}}

\begin{document}

\title{Multiple pendulum and nonuniform distribution of average kinetic energy}

\author{Tetsuro Konishi}
\email[e-mail: ]{tkonishi@isc.chubu.ac.jp}
\affiliation{College of Engineering, Chubu University, Kasugai 487-8501, Japan.}

\author{Tatsuo Yanagita}
\affiliation{Department of Engineering Science, Osaka Electro-Communication University, Neyagawa 572-8530, Japan}

\date{\today}

\begin{abstract}
 Multiple pendulums are investigated numerically and analytically to clarify the nonuniformity of average kinetic  energies of particles.
The nonuniformity is attributed to the system having constraints and it is 
consistent with the generalized principle of the equipartition of energy.
With the use of explicit expression for Hamiltonian of a multiple pendulum, approximate expressions for temporal and statistical average of kinetic energies are obtained, where the average energies are expressed in terms of masses of particles.
In a typical case, the average kinetic energy is large for particles near the end of the pendulum and small for those near the root.
Moreover, the exact analytic expressions for the average kinetic energy of the particles are obtained for a double pendulum. 
\end{abstract}


\maketitle

\section{Introduction}
Pendulums are useful to explore the fundamental behavior of various physical
systems. A simple pendulum is a good example of a system that can demonstrate 
periodic motion~\cite{galileo}.
When the amplitude of the oscillation is small, i.e., the total energy is small,
the periodic motion of a simple pendulum is well approximated by that of a harmonic
oscillator, wherein the period of the oscillation does not depend on the amplitude~\cite{goldstein}.
  
The fundamental modes and principle of superposition can be understood by studying a double pendulum~\cite{Daniel-Bernoulli-pendulum-1,Daniel-Bernoulli-pendulum-2,Cannon-the-evolution-of-dynamics,goldstein}.
When the amplitude of oscillation is small, there are two special motions
called the fundamental modes, wherein both the upper and lower pendulums oscillate in the same period.
For a small-amplitude oscillation, one can understand that every motion of the double pendulum is expressed by the superposition of the two fundamental modes.
The principle of superposition is a key concept for understanding various linear phenomena. 
Since both fundamental modes of the double pendulum indicates periodic motion, the double pendulum exhibits periodic or quasiperiodic motion when the amplitudes are small.

Meanwhile, a double pendulum is also a good example of a system that indicates chaotic motion~\cite{Lichtenberg-Lieberman,ott-chaos-text,tabor-book,doi:10.1119/1.16860,STACHOWIAK2006417,doi:10.1119/1.16860,doi:10.1119/1.3052072,HolgerDullin-DoublePendulum-Melnikov-1994,STACHOWIAK20153017,Ivanov-DoublePendulum-I-1999,Ivanov_2001,Ivanov-DoublePendulum-III-2000,Ivanov-DoublePendulum-IV-2001}. If a large energy is assigned and the amplitudes of displacements
are not small, the motion of a double pendulum is no longer regular, and 
chaotic behavior is observed. 
Multiple pendulums with three or more degrees of freedom~\cite{Daniel-Bernoulli-pendulum-1,Daniel-Bernoulli-pendulum-2,Cannon-the-evolution-of-dynamics} also exhibit chaotic behavior~\cite{yanagita-gakkai-1,yanagita-gakkai-2,yanagita-gakkai-3}.

When chaos is strong in a multiple pendulum, one can expect that
it admits a statistical description; the long time average of the physical quantity
is considered to be approximately equal to the average over the energy surface.
 Then, each particle in the pendulum can be
considered as a subsystem connected to a heat bath that comprises the rest of the
pendulum; then, the particles are in a thermal equilibrium. 

One may expect that the sytem in thermal equilibrium is almost uniform. However, a careful observation of the motions of a multiple pendulum %
indicate that the particle at the end of the pendulum moves
faster than the particle nearest to the root of the 
pendulum, even if the masses of all particles are the same.
In fact, one author conducted numerical computations and revealed that the time average of 
the kinetic energies of a multiple pendulum are different~\cite{yanagita-gakkai-1,yanagita-gakkai-2,yanagita-gakkai-3}.
In this paper, we revisit the observation and explain the origin of
differences in the average kinetic energies using the generalized principle
of the equipartition of energy~\cite{Tolman-PR-1918,Tolman-book,kubo-book,chain-letter-JSTAT-2009}. The nonuniformity of the average kinetic energy is consistent
with the generalized principle of equipartition of energy.
Thus, the average kinetic energy can be nonuniform under a thermal equilibrium.

The remainder of this paper is organized as follows. In Section \ref{sec:model}, we describe our model,  a multiple pendulum. Then, we provide an explicit form of the Lagrangian and Hamiltonian of a multiple pendulum.
In Section \ref{sec:numerical}, we present the results of numerical computation where the values of 
time-average kinetic energies of particles in the multiple pendulum are not equal, at the same time the generalized principle of equipartition holds.
In Section \ref{sec:analytic}, we explain the nonuniform distribution of average kinetic energy 
based on statistical mechanics; further, an exact expression is obtained for the double pendulum. Finally, Section V presents the summary and the discussions.

\section{Model}\label{sec:model}
\begin{figure}[t]
  \centering
  \includegraphics[width=\hsize]{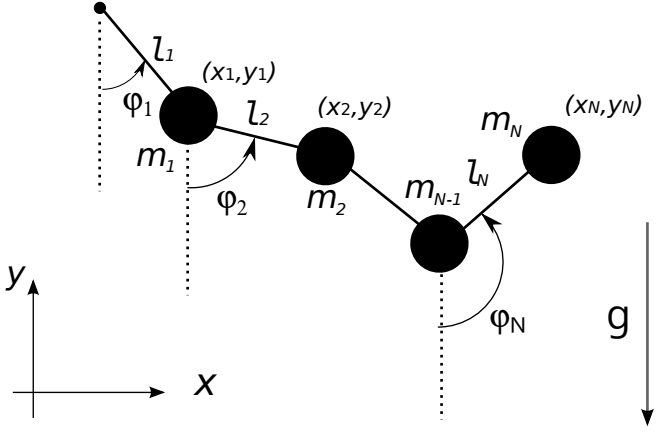}
  \caption{Multiple pendulum: Definition of variables and parameters}
  \label{fig:npend-def}
\end{figure}

The multiple pendulum examined in this study is composed of $N$ particles serially connected by $N$ massless links of fixed length in a uniform gravitational field, as illustrated in Fig.\ref{fig:npend-def}.
One end is fixed to a point we define as the origin $(0,0)$.
The particles and links pass through each other if they arrive at the same position.
The particles move in a fixed vertical plane, which is defined as the $xy$-plane.
The $x$-axis is considered to be in the horizontal direction, and the $y$-axis, in the vertical direction with the upward direction representing the 
positive $y$ direction.
We let $m_i$, $\VEC{r_i}\equiv (x_i, y_i)$, and $\ell_i$ 
represent the mass of the $i$'th particle, position of the $i$'th particle,
and length of $i$'th link, respectively.
$g$ represents the constant of gravitational acceleration.
The distance between $i$'th and $i+1$'th particles 
is fixed for all $i$. In this sense, the system has constraints.

In terms of the Cartesian coordinates $(x_i,y_i)$, the kinetic energy of the $i$'th particle
$K_i$, total kinetic energy $K$, and potential energy $U$ are 
\begin{align}
K_i& = \frac{m_i}{2}(\dot x_i^2+\dot y_i^2)  \label{eq:ke-linear-i} \ , \\
K &=\sum_{i=1}^N K_i \ , \\
U &= \sum_{i=1}^N m_i g y_i \  . \label{eq:potential}  
\end{align}

Using these equations, the system is defined by a Lagrangian $L$ and constraints $G_i$, which are given as
\begin{align}
  \label{eq:lag-npend}
  L &=K-U= \sum_{i=1}^N \frac{m_i}{2}\left( \dot x_i^2 + \dot  y_i^2\right)
  - \sum_{i=1}^N m_i g y_i \ , \\
  G_i &\equiv \left| \VEC{r_{i}} - \VEC{r_{i-1}}\right|^2 - \ell_i^2
     = 0 \ , i=1,2,\cdots, N ,
    \label{eq:constraint-npend}
\end{align}
where $\VEC{r_0}\equiv (x_0,y_0) \equiv (0,0)$ and a dot over each symbol represents derivative with respect to time  e.g., $\dot x_i = \frac{dx}{dt}$.

If we denote $\varphi_i$ as the angle between the $i$'th link and the $-y$ direction
(direction of gravity), we have
\begin{align}
x_i&= x_{i-1} + \ell_i \sin\varphi_i \ \
  = \sum_{j=1}^i \ell_j \sin\varphi_j   \label{eq:x-by-varphi},
 \\
  y_i&= y_{i-1} - \ell_i \cos\varphi_i  \ \  
  = -\sum_{j=1}^i \ell_j \cos\varphi_j \, .
\label{eq:y-by-varphi}
\end{align}

The Lagrangian  (\ref{eq:lag-npend}) 
is then expressed in terms of $\varphi_i$, $i=1,2,\cdots, N$ as 
  \begin{align}
    L &= K - U  \ , %
    \\
    K &=\sum_{i,j=1}^N\frac{1}{2}A(\varphi)_{ij}\dot\varphi_i\dot\varphi_j  \ , 
    \label{eq:total-ke}\\
    U &= -\sum_{i=1}^N m_i g\sum_{j=1}^i  \ell_j\cos\varphi_j  \ , 
    \label{eq:potential-energy}
  \end{align}
where the $N\times N$ matrix $A$ is defined as 
\begin{equation}
  A(\varphi)_{nk}\equiv
\left(  \sum_{i=\max(n,k)}^N m_i\right)
\ell_n\ell_k
\cos(\varphi_n -\varphi_k) \label{eq:a-ij} \ \ . 
\end{equation}
The derivations of Eqs. (\ref{eq:total-ke}) and (\ref{eq:a-ij}) are presented in Appendix ~\ref{sec:lagrangian-general}.

Using Eqs.(\ref{eq:x-by-varphi}) and (\ref{eq:y-by-varphi}),  $K_i$ can be expressed in the quadratic form of $\dot\varphi$ as 
\begin{align}
  K_i& = \frac{m_i}{2}\sum_{j=1}^i\sum_{k=1}^i \dot\varphi_j\dot\varphi_k \ell_j\ell_k \cos(\varphi_j-\varphi_k)\label{eq:Ki-in-dot-varphi} \\
& = \frac{1}{2}\sum_{j,k}A_{jk}^{(i)}\dot\varphi_j \dot \varphi_k 
  \label{eq:Ki-linear-dotvarphi-Ai}
\end{align}
where $A^{(i)}$ represents a $N\times N$ matrix defined as
\begin{equation}
  A_{jk}^{(i)} = 
  \begin{cases}
m_i\ell_j\ell_k \cos(\varphi_j-\varphi_k) & \cdots j\le i \  \text{and}  \ k \le i \ , \\
    0  & \cdots \text{otherwise} \ .
  \end{cases}
\end{equation}
Because $\sum_i K_i = K$, the sum of $A^{(i)}$ with respect to $i$ is equal to the matrix $A$:
\begin{equation}
  \sum_{i=1}^NA_{jk}^{(i)} = A_{jk} \ .
  \label{eq:mat-A-i}
\end{equation}
Matrices $A$ and $A^{(i)}$ depend on coordinates $\varphi$.

The momentum $p_i$ canonically conjugate to the coordinate $\varphi_i$  is defined as
\begin{equation}
  \label{eq:momentum}
     p_i\equiv\pdd{L}{\dot\varphi_i} \ .
\end{equation}
Then, we have
\begin{align}
  p_n&
= \sum_{k=1}^N A_{nk}\dot\varphi_k 
\,,
\label{eq:can-mom}\\
K&=\frac{1}{2}\sum_{j,k}p_jA^{-1}_{jk}p_k  \ .
\label{eq:ke-in-p}
\end{align}

$K_i$ using momentum $p$ is expressed as
\begin{equation}
  K_i %
  =\frac{1}{2}\sum_{j,k,\xi,\eta}A_{jk}^{(i)}A^{-1}_{j\xi}A^{-1}_{k\eta}p_{\xi}p_{\eta}\ , 
  \label{eq:Ki-linear-pi}
\end{equation}

Using Eq.(\ref{eq:ke-in-p}), the Hamiltonian of a multiple pendulum is given as
\begin{equation}
  H = \frac{1}{2}\sum_{j,k=1}^Np_jA^{-1}_{jk}p_k   -\sum_{i=1}^N m_i g\sum_{j=1}^i  \ell_j\cos\varphi_j \ .
  \label{eq:Ham-npend}
\end{equation}

\section{Numerical Example}\label{sec:numerical}
\subsection{Simulation Method}

First, let us explain the methods of numerical simulation 
using which we integrate the equation of motion of the multiple pendulum.
The equation of motion derived from the Lagrangian written in terms of
$\varphi$
is complicated because of the 
dependence of the kinetic energy on $\varphi$. Hence, we use the equation of motion
written in terms of the Cartesian coordinates $xy$ derived from Eqs.(\ref{eq:lag-npend})
and (\ref{eq:constraint-npend}). The equation of motion includes the terms from the  
constraint characterized by the coefficients called ``Lagrange multipliers''~\cite{goldstein}.
We numerically evaluate the Lagrange multipliers at each step of the integration
to ensure that the constraints in Eq.(\ref{eq:constraint-npend}) are satisfied. The idea used in this method is the same as that in the ``RATTLE'' algorithm~\cite{leimkurler-reich,CHARMM,Vesely-DoublePendulum-AJP-2013}.
We use the fourth-order symplectic integrator composed of three second-order symplectic integrators ~\cite{leimkurler-reich,SI-yoshida} incorporating forces from the constraints.

We calculate the kinetic energies of the particles of the multiple pendulum.
Let us denote $K_i(t)$ as the kinetic energy of the $i$'th particle (Eq.(\ref{eq:ke-linear-i})) at time $t$.
The time average of $K_i(t)$ is defined as
\begin{equation}
  \label{eq:ke-linear-lt-ave}
  \overline{K_i}(t) \equiv
  \frac{1}{t}\int_0^{t}K_i(t')\,dt'
 \  . %
\end{equation}

\subsection{Simulation Results}
Let us observe some typical time evolutions of the multiple pendulum obtained from 
the numerical simulation.
Here, we adopt the initial condition $\dot x_i(0) = \dot y_i(0)=0$ for all $i$;
$\varphi_i(0) = \theta$ for all $i$. This is a stretched configuration
with angle $\theta$.

\begin{figure}[htbp]
  \includegraphics[width=0.9\hsize]{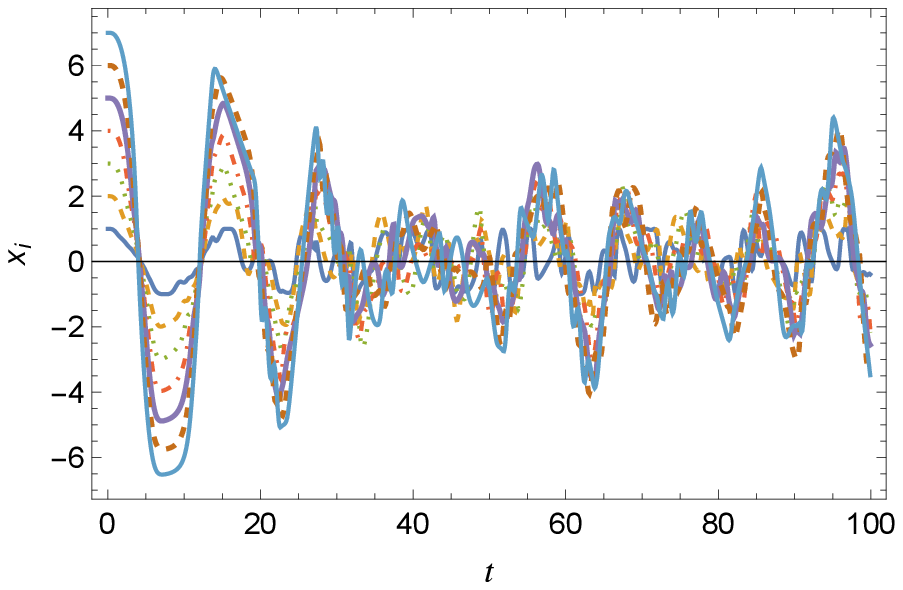}
  \includegraphics[width=0.9\hsize]{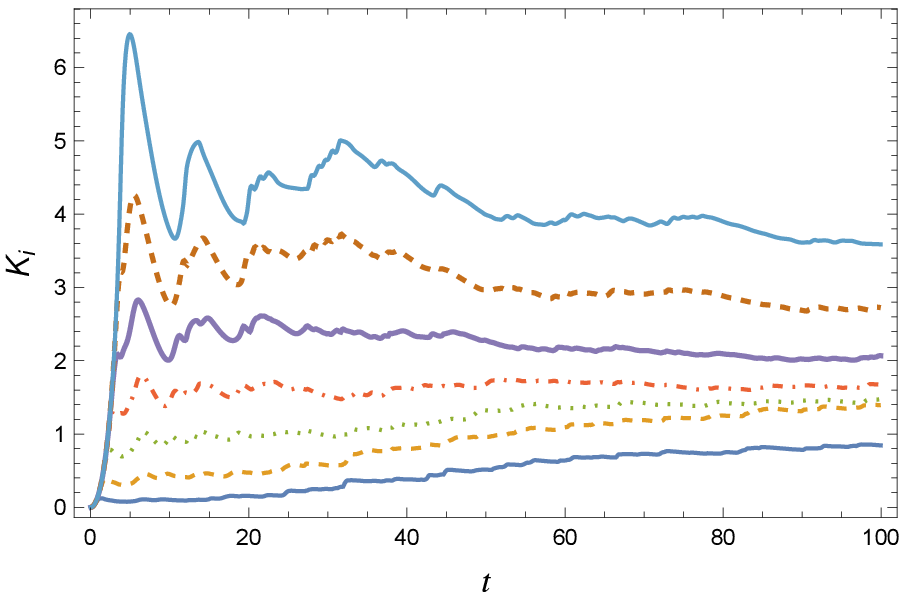}
  \caption{(top): Time evolution of a multiple pendulum. The horizontal axis represents time, and the vertical axis represents the $x$ coordinate of each particle $x_i(t)$.  $N = 7$, $m_i = 1$, and $\ell_i = 1$ for all $i$; further, $g = 1$. The time step for the numerical integration is $\Delta t = 0.001$. The  initial conditions are  $\varphi_i = \pi/2$ and $\dot\varphi_i = 0$ for all $i$.  
    (bottom): Average kinetic energies $\overline{K_i}(t)$ vs. $t$ calculated from 
the time evolution shown in the top panel. The lines represent $\overline{K_1}(t)$, 
$\overline{K_2}(t)$,  $\cdots$, and $\overline{K_7}(t)$, from the bottom to the top. 
  }
  \label{fig:initial-theta-0.5pi}
\end{figure}

Let us consider the initial angle as $\theta = \pi/2$.
The top panel of Fig.\ref{fig:initial-theta-0.5pi} shows the chaotic motion after a short transient;
the bottom panel shows the temporal evolution of the average kinetic energies $\overline{K_i}(t)$. Further, $\overline{K_i}(t)$ does not converge to 
a single value; on the contrary, they converge to different values, i.e., the average kinetic energy is not uniform.

\begin{figure}[htbp]
  \centering
 \includegraphics[width=0.8\hsize]{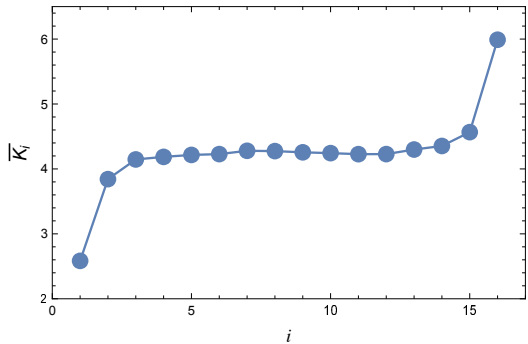}
 \caption{Long time average of  kinetic energy  $\overline{K_i}(T)$ vs. $i$.
   $N = 16$,  $m_i = 1$, and $\ell_i = 1$ for all $i$; further, $g = 1$.
The initial condition is $\varphi_i = \pi/2$, $\dot\varphi_i = 0$ for all $i$.
$T = 1.0\times 10^4$.
}
  \label{fig:ke-linear}
\end{figure}

The nonuniformity in the $\overline{K_i}$ remains
even if we take a longer time for averaging.
Fig.~\ref{fig:ke-linear} shows a plot of the average kinetic energy of each particle
$\overline{K_i}$ (Eq.(\ref{eq:ke-linear-lt-ave})) against $i$ for $N = 16$.
The values of  $\overline{K_i}$ are not the same.
The $\overline{K_i}$ of particles near the 
end are large, whereas those of particles near the root are small.
This is consistent with Yanagita et al.'s first observation~\cite{yanagita-gakkai-1,yanagita-gakkai-2,yanagita-gakkai-3}. 

\subsection{Non-uniformity of Average Kinetic Energy and  the Generalized Principle of the Equipartition of Energy}
Let us consider the relation between the non-uniformity of $\overline{K_i}$ we observed in the previous subsection and the generalized principle of the equipartition of energy~\cite{Tolman-PR-1918,Tolman-book,kubo-book,chain-letter-JSTAT-2009}.

It is natural to assume that
the set of points $\left(\varphi_i(t), p_i(t)\right), i = 1,2,\cdots, N$ generated from the time series is close to the microcanonical ensemble because the motion of the multiple pendulum considered in this study is highly chaotic. 
Further, each single particle in the multiple pendulum can be regarded as a subsystem 
attached to a ``heat bath'' composed of the rest of the pendulum ($N-1$ particles),
and we can roughly approximate the statistical distribution of the state of the single particle
as a canonical distribution at some temperature.
This assumption allows us to interpret the original phenomena of the nonuniformity of the temporal 
average $\overline{K_i}$ as the nonuniformity of the thermal average $\left\langle K_i \right\rangle$ .
Below, we explain the nonuniformity of
the thermal average using the generalized principle of the equipartition of energy.

Eq.(\ref{eq:total-ke}) shows that the kinetic energy of the system has off-diagonal elements that depend on the coordinate $\varphi$.
Thus, we cannot apply the ordinary form of the principle of the equipartition
of energy. Instead, we use the generalized principle of the equipartition
of energy, which states 
\begin{equation}
\left< K_i^{(c)}\right> = \frac{1}{2}k_B T \ . 
\label{eq:generalized-principle-of-equipartition}
\end{equation} 
Here, $T$ represents the temperature, $k_B$ denotes Boltzmann's constant, 
the bracket $\left\langle \cdots \right\rangle$ represents the thermal average, 
and $K_i^{(c)}$ is defined as
\begin{equation}
 K_i^{(c)} \equiv \frac{1}{2}p_i \pdd{K}{p_i},
\label{eq:ke-canonical}
\end{equation}
where $p_i$ represents the momentum canonically conjugate to the coordinate $\varphi_i$.
In Eq.(\ref{eq:ke-canonical}), the summation with respect to $i$ is not considered.
We call $K_i$ and $ K_i^{(c)}$ the ``linear kinetic energy'' and ``canonical kinetic energy”, respectively. 

For the multiple pendulum, $K_i^{(c)}$~(Eq.(\ref{eq:ke-canonical})) is
\begin{equation}
  K_i^{(c)} =  \sum_{k=1}^N\frac{1}{2}p_i A^{-1}_{ik}p_k \, .
  \label{eq:ke-canonical-i}
\end{equation}
This is different from the kinetic energy of the $i$'th particle $K_i$
defined in (Eq.(\ref{eq:Ki-linear-pi})), hence 
\begin{equation}
  \label{eq:lke-cke-different}
  K_i\ne K_i^{(c)} \ .
\end{equation}
Using Eqs.~(\ref{eq:generalized-principle-of-equipartition}) and (\ref{eq:lke-cke-different}), $\left\langle K_i \right\rangle$, the average of  $K_i$, does not take the same value at the thermal equilibrium in general. 

\begin{figure}%
  \centering
  \includegraphics[width=8cm]{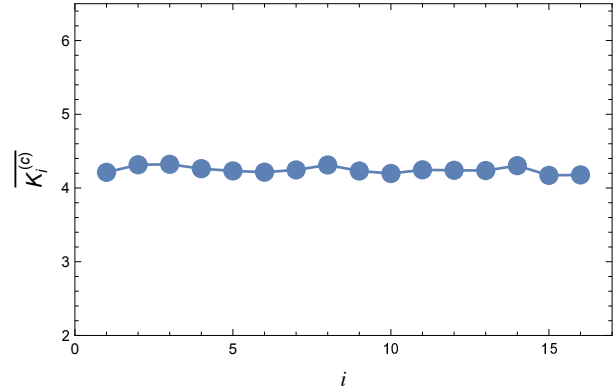}
  \caption{
    Long time average of canonical kinetic energy
    $\overline{K_i^{(c)}}$ (Eq.(\ref{eq:ke-canonical})) vs. $i$.
    This plot was obtained using the same time series as that in  Fig. \ref{fig:ke-linear}.
  }
  \label{fig:ke-canonical}
\end{figure}

Fig.~\ref{fig:ke-canonical} shows the time average of  canonical kinetic energy ~Eq.(\ref{eq:ke-canonical})
of the multiple pendulum using the same time series as in Fig.~\ref{fig:ke-linear}.
Here, we can see that $\overline{K_i^{(c)}}$ 
takes almost the same value and the generalized principle of the equipartition of energy is realized.
That is, even when the thermal equilibrium is established and the generalized principle of equipartition of energy holds, the average of $K_i$ takes different values and the nonuniformity of the average kinetic energy is realized in the thermal equilibrium.

\section{Analytical explanation}\label{sec:analytic}

\subsection{Temporal average}
If we denote
\begin{equation}
  \Delta x_i \equiv x_i - x_{i-1} \ ,    \
  \Delta  y_i \equiv  y_i - y_{i-1} \ ,  
\end{equation}
Eq.(\ref{eq:ke-canonical-i}), the ``canonical kinetic energy'' of the $i$'th degree of freedom,  is expressed as
\begin{equation}
  K_i^{c}
= 
\sum_{j=1}^N\frac{1}{2}\left(\sum_{n=\max(i,j)}^N m_n\right)
\left[
\Delta \dot x_i \Delta \dot x_j
+ 
\Delta \dot y_i \Delta \dot y_j
\right] \ \ .
\label{eq:ke-can-i-2}
\end{equation}
where \ $\Delta \dot x_1 = \dot x_1$ \  and \ 
$\Delta \dot y_1 = \dot y_1$ \ 
because $x_0 \equiv y_0 \equiv 0$.
Detail of calculation is shown in Appendix~\ref{sec:app-canonical-ke}.

Let us assume
\begin{equation}
\overline{\Delta \dot x_i \Delta \dot x_j}  = 0  \ , \ 
\overline{\Delta \dot y_i \Delta \dot y_j}  = 0  \ 
\ ( i \ne j) \ 
. 
\label{eq:npend-approx}
\end{equation}
This assumption implies that each link in the multiple pendulum
rotates statistically independently.

Then, we have by straightforward calculation that
\begin{align}
  \overline{K_i^{c}}
&=
\frac{1}{2}\left(\sum_{n=i}^N m_n\right)
\left[
\left(\overline{\dot x_i^2}+ \overline{\dot y_i^2} \right)
 - \left( \overline{\dot x_{i-1}^2}
+ 
 \overline{\dot y_{i-1}^2}\right)
\right]  \\
&=\left(\sum_{n=i}^N m_n\right) 
\left( \frac{1}{m_i} \overline{K_i}-\frac{1}{m_{i-1}}
  \overline{K_{i-1}}
\right) \ .
\label{eq:ke-can-i-ave-2}
\end{align}

Assuming that the temporal average $\overline{K^c_i}$  is equal to the thermal average 
$\left<K^c_i\right>$
and 
using the generalized principle of equipartition of energy~ \cite{Tolman-PR-1918,Tolman-book,kubo-book} for the ``canonical kinetic energy'' we have
$\overline{K^c_i} = \frac{1}{2}k_B T$,
and the long time average of ``linear kinetic energy'' of
the $i$'th particle $K_i$ is recursively expressed as
\begin{equation}
\overline{ K_i} = \frac{m_i}{m_{i-1}}\,\overline{K_{i-1}}
+\frac{m_i}{\sum_{n=i}^N m_n}\,\cdot\frac{1}{2}k_BT \ .
\label{eq:ke-linear-time-ave}
\end{equation}
Since $(x_0,y_0)=(0,0)$ and $K_0\equiv 0$, we have
\begin{equation}
  \overline{ K_i} = m_i\left(\sum_{j=1}^{i}\frac{1}{\sum_{n=i}^N m_n}\right)\cdot \frac{1}{2}k_B T \, .
\label{eq:ke-linear-time-ave-2}
\end{equation}
Hence, the temporal average of the kinetic energy is nonuniform.

If all masses have the same value
$ m_1 = m_2 = \cdots = m_N$,
then $\overline{ K_n}$'s are explicitly expressed as
\begin{equation}
 \overline{K_i} = \left(\sum_{j=1}^i\frac{1}{N-j+1}\right)\frac{k_B T}{2} 
\label{eq:npend-non-equipart-1}
\end{equation}
so that the average linear kinetic energies $\overline{K_i}$ 
can monotonically increase from the root to the end of the pendulum as
\begin{equation}
\overline{K_1} < \overline{K_2} < \cdots < \overline{K_N}  \ .
\label{eq:npend-non-equipart-2}
\end{equation}
This is consistent with the numerical results shown in Fig.~\ref{fig:ke-linear}.

\subsection{Statistical Average}
The nonuniformity of the average energy can be explained using a statistical method.
Suppose the system is under thermal equilibrium at 
temperature $T$. The statistical average of the linear 
kinetic energy $K_i$ is then defined as
\begin{equation}
  \left<K_i\right> \equiv \frac{1}{Z}\int K_i e^{-\beta H} d\Gamma  \, ,
  \label{eq:thermal-ave-ki-linear}
\end{equation}
where $Z\equiv \int  e^{-\beta H} d\Gamma $,
$ \beta \equiv 1/k_B T$, 
and 
$  d\Gamma \equiv d^N\! p\,d^N\! \varphi$ \ . 

Using the expression for $K_i$, 
\begin{equation}
    \left<K_i\right> 
=
\frac{m_i}{2}\sum_{j=1}^i\sum_{k=1}^i
\ell_j \ell_k
\left<
\dot\varphi_j \dot \varphi_k \cos \varphi_{jk} 
\right>
\ , 
\label{eq:ave-Ki-linear-dotvarphi}
\end{equation}
where $\varphi_{jk} \equiv \varphi_j - \varphi_k $ .

Using Eq.(\ref{eq:Ki-linear-pi}), we can express Eq.(\ref{eq:ave-Ki-linear-dotvarphi})
 in terms of the canonical momenta $p_i$ as
 \begin{equation}
    \left\langle K_i\right\rangle 
=
\frac{1}{2}\sum_{j,k,\xi,\eta}
\left\langle
A_{jk}^{(i)}A^{-1}_{j\xi}A^{-1}_{k\eta}p_{\xi}p_{\eta}
\right\rangle \ .
   \label{eq:ave-Ki-linear-pi-1}
 \end{equation}

Integrating by parts with respect to $p$ yields
\begin{align}
    \left\langle K_i\right\rangle 
&= \frac{1}{2\beta}
    \left\langle
      \mbox{tr}\left( A^{(i)}A^{-1}\right) 
    \right\rangle 
  \label{eq:ave-Ki-linear-pi-trace}
\\
&=
\frac{1}{2\beta}\frac{1}{Z}\int     \mbox{tr}\left( A^{(i)}A^{-1}\right) 
e^{-\beta 
\left(
  \frac{1}{2}pA^{-1}p + U(\varphi)
\right)}
 d\Gamma \ .
  \label{eq:ave-Ki-linear-pi-trace-int}
\end{align}
Here, matrices $A^{-1}$ and $A^{(i)}$ depend on $\varphi$.

In Eq.(\ref{eq:ave-Ki-linear-pi-trace-int}),  we  first perform  integration with respect to $p$,  which is 
 a multidimensional Gaussian integral. The result is
\begin{equation}
  \int e^{-\beta 
\left(
  \frac{1}{2}pA^{-1}p
\right)} d^N p
=\left(\frac{2\pi }{\beta}\right)^{N/2} \sqrt{\det A} \ .
\end{equation}
Hence, we have
\begin{align}
  Z &= \left(\frac{2\pi }{\beta}\right)^{N/2}\int \sqrt{\det A}\, e^{-\beta U(\varphi)}d^N\varphi
      \ , \\
    \left\langle K_i\right\rangle 
&= \frac{1}{2\beta}
\frac{  \int
      \mbox{tr}\left( A^{(i)}A^{-1}\right) 
\sqrt{\det A} \, e^{-\beta U(\varphi)}d^N\varphi
}
{\int \sqrt{\det A}\, e^{-\beta U(\varphi)}d^N\varphi} \ .
\label{eq:ave-Ki-linear-pi-trace-varphi}
\end{align}
In the following, we use these equations to express $\left\langle K_i\right\rangle$ for multiple and double pendulums.

\subsubsection{Multiple Pendulum with an arbitrary number of particles}
Let us adopt ``diagonal approximation''
\begin{equation}
\left<
\dot\varphi_j \dot \varphi_k \cos \varphi_{jk}  
\right>
= 0 \ \text{for} \ j \ne k \, ,
\label{eq:diagonal-approx-varphi}
\end{equation}
and
\begin{equation}
 \left< A^{-1}_{jj}\right>\approx \left< \frac{1}{A_{jj}} \right>
  = \frac{1}{\left(\sum_{i=j}^N m_i\right)\ell_j^2 } \ .
\end{equation}
These approximations assumes that each link rotates statistically independently,
and we omit all non-diagonal elements of matrix $A$, where phase factors
such as $\cos(\varphi_i-\varphi_j)$ are included.

Then, we obtain 
\begin{equation}
\left< K_i \right>
=
m_i\left(\sum_{j=1}^i \frac{1}{\sum_{n=j}^N m_n}\right)\cdot 
\frac{1}{2}k_B T \,    .
\label{eq:ke-linear-tave-1}
\end{equation}
This expression is equivalent to that in Eq.(\ref{eq:ke-linear-time-ave-2}) if the thermal average on the left-hand side $\left<\cdots\right>$ is replaced by the time average $\overline{\cdots}$.
The details of the calculation are summarized in Appendix \ref{sec:Ki-ave-approx-stat}.

This result indicates that when all masses are the same,
the average linear kinetic energies are monotonically increasing
from the root to the end of the pendulum, as shown in Fig.~\ref{fig:ke-linear}.
\begin{equation}
\left<K_1\right> < \left<{K_2}\right> < \cdots < \left< {K_N}\right>  \ .
\label{eq:npend-non-equipart-thermal}
\end{equation}
\subsubsection{Double pendulum}
\begin{figure}
  \centering
  \includegraphics[width=0.8\hsize]{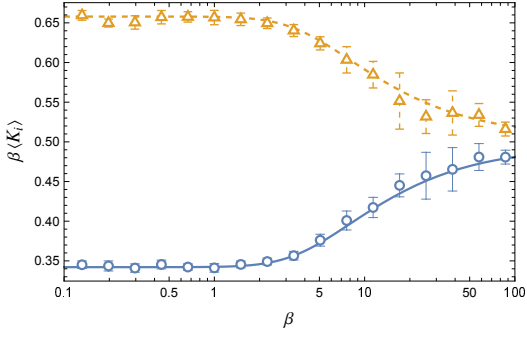}
  \caption{$\beta$-dependence of $\left\langle K_i\right\rangle$ for double pendulum.
    $\beta \cdot \left\langle K_i\right\rangle$ are plotted.
    The upper and lower curves represent $\left\langle K_2\right\rangle$ and $\left\langle K_1\right\rangle $, respectively, obtained from Eqs.(\ref{eq:ave-K2-result}) and (\ref{eq:ave-K1-result}), respectively. The symbols represent the numerically obtained values of Eq.(\ref{eq:thermal-ave-ki-linear}) using the Markov chain Monte Carlo method.  $m_i = 1/2$, $\ell_i = 1$ for $i = 1,2$. $g=1$.}
  \label{fig:Ki-ave-exact-and-mcmc}
\end{figure}

For the case of the double pendulum, i.e., for the $N = 2$ case, we can obtain the exact expressions for
$\left\langle K_i \right\rangle$.

For $N = 2$, matrices $A$, $A^{(i)}$, 
defined by Eqs. (\ref{eq:a-ij}) and  (\ref{eq:mat-A-i}), respectively, 
and  $\det A$ reads
\begin{align}
  A &=M
  \begin{pmatrix}
    \ell_1^2 &   \mu_2\ell_1\ell_2 C_{12} \\
\mu_2  \ell_1\ell_2C_{12}
& \mu_2\ell_2^2
  \end{pmatrix} \ , 
\label{eq:mat-A-dp}
\\
A^{(1)} 
&=
M             \mu_1 \ell_1^2         \begin{pmatrix}
1 & 0 \\ 0 & 0
          \end{pmatrix} \ , 
\label{eq:mat-A-1-dp}
\\
A^{(2)}&=
M \mu_2        \begin{pmatrix}
          \ell_1^2 &   \ell_1\ell_2 C_{12} \\
  \ell_1\ell_2C_{12}
& \ell_2^2
         \end{pmatrix} \ ,
  \label{eq:mat-A-2-dp}
\\
\det A
&=
M^2\mu_2 \ell_1^2\ell_2^2
\left( 1-\mu_2 C_{12}^2\right) \ , 
\label{eq:detA}
\end{align}
where
\begin{align}
  C_{12} &= \cos(\varphi_2 - \varphi_1) \ , \\
  M &= m_1+m_2 \ , \\
  \mu_i&=m_i/M \ .
\end{align}
From these, we obtain
\begin{align}
   \mbox{tr}\left( A^{(1)}A^{-1}\right) 
&=\frac{\mu_1}{ 1-\mu_2 C_{12}^2} \ , 
  \label{eq:trA1Ainv-dp}
\\
   \mbox{tr}\left( A^{(2)}A^{-1}\right) 
&=
\frac{(1+\mu_2)-2\mu_2C_{12}^2}{
 1-\mu_2 C_{12}^2
}
= 2-\mbox{tr}\left(A^{(1)}A^{-1}\right)\ .
  \label{eq:trA2Ainv-dp}
\end{align}
Here, we used $\mu_1+\mu_2 = 1 $.

We obtain
the exact expressions of $\left\langle K_i \right\rangle$ 
for a double pendulum by substituting these into Eq.(\ref{eq:ave-Ki-linear-pi-trace-varphi}) 
and expanding $\det A$ as a series of $\mu_2 C_{12}^2$.
\begin{align}
\left\langle K_1 \right\rangle
&=  
\frac{1}{2\beta}
\frac{\displaystyle 
\mu_1
\sum_{n=0}^\infty \frac{(2n-1)!!}{n! 2^n}\mu_2^n
R_n(\alpha_1,\alpha_2) 
}
{\displaystyle
I_0(\alpha_1)I_0(\alpha_2) 
 -\sum_{n=1}^\infty  \frac{(2n-3)!!}{n! 2^n}
 \mu_2^n R_n(\alpha_1,\alpha_2) 
} \ , 
\label{eq:ave-K1-result}
\\
\left\langle K_2 \right\rangle
&=
\frac{1}{\beta}-\left\langle K_1 \right\rangle \ ,
\label{eq:ave-K2-result}
\end{align}
where
\begin{align}
R_n(\alpha_1,\alpha_2)
&=
\sum_{j=0}^{n}{}_{2n}C_{2j}
   \left\{(2(n-j)-1)!!\right\}^2
\nonumber\\
& \cdot  \left\{
   \frac{\partial^{2j}}{\partial \alpha_1^{2j}}
   \left(
   \frac{I_{n-j}(\alpha_1)}{\alpha_1^{n-j}} 
   \right)
   \right\}
   \left\{
   \frac{\partial^{2j}}{\partial \alpha_2^{2j}}
   \left(
   \frac{I_{n-j}(\alpha_2)}{\alpha_2^{n-j}} 
   \right)
   \right\}
\\
  \alpha_1 &=\beta (m_1+m_2)g\ell_1 , \label{eq:alpha-g}\\ %
  \alpha_2 &=\beta m_2 g\ell_2 , \label{eq:alpha-ell}  %
\end{align}
and $I_n(z)$ is the modified Bessel function of the $n$'th order~\cite{DLMF}.

Fig.~\ref{fig:Ki-ave-exact-and-mcmc} shows the values of $\left\langle K_i\right\rangle$ , $i = 1,2$, calculated from Eqs.(\ref{eq:ave-K1-result}) and (\ref{eq:ave-K2-result}). We show the result of Eq.(\ref{eq:thermal-ave-ki-linear}) obtained from Markov Chain Monte Carlo method by symbols to check the validity of Eqs.(\ref{eq:ave-K1-result}) and (\ref{eq:ave-K2-result}). We see that both calculations agree quite well.

\begin{figure*}
  \centering
\includegraphics[width=0.3\hsize]{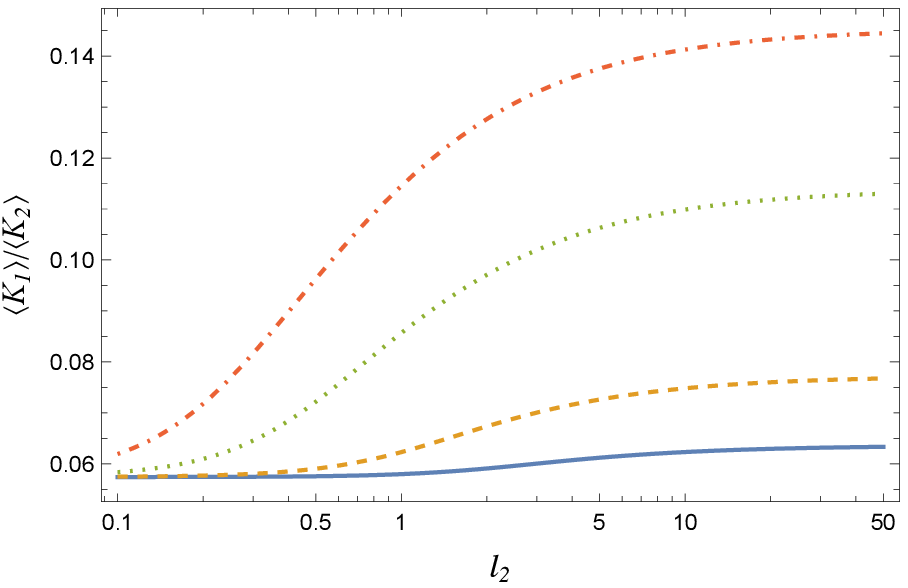}  
\includegraphics[width=0.3\hsize]{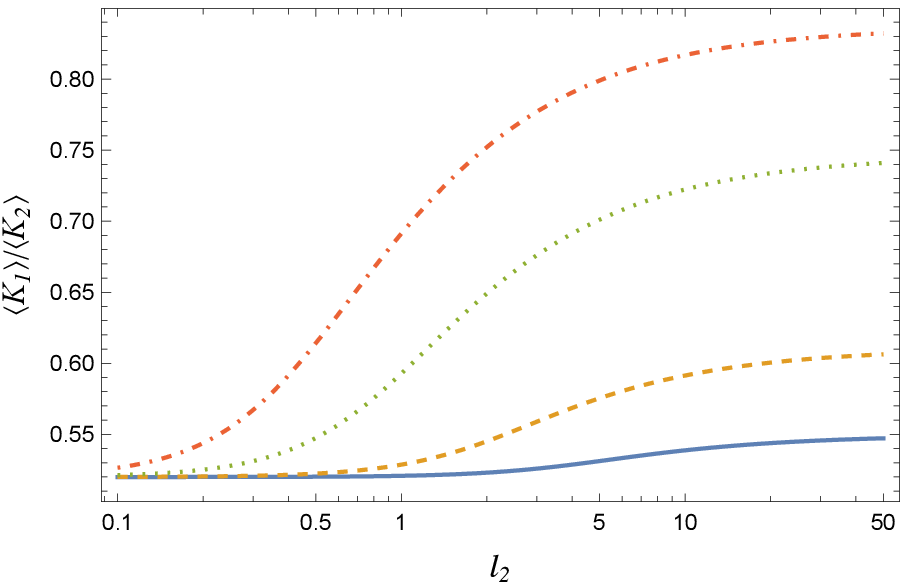}  
\includegraphics[width=0.3\hsize]{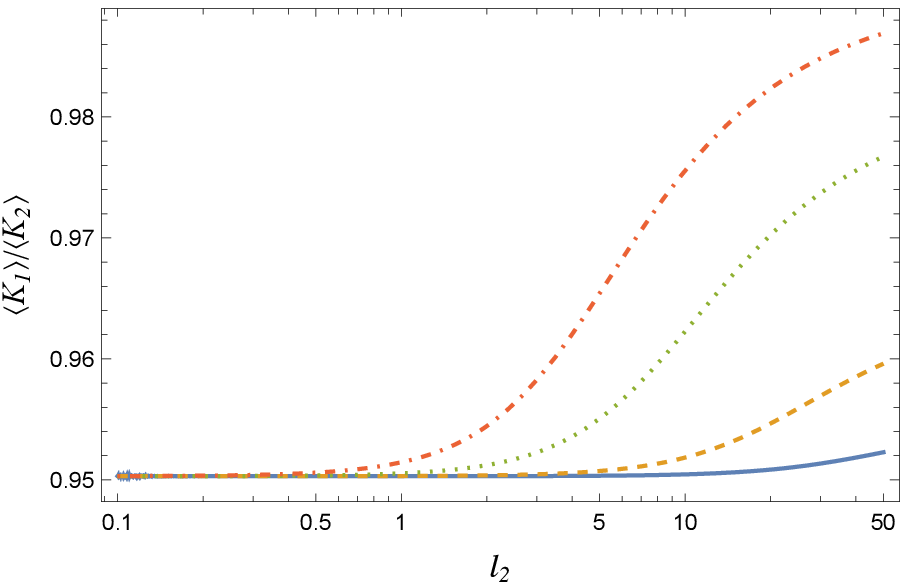}  
  \caption[$K_1/K_2$ vs. $\ell_2$]{$\ell_2$ dependence of the ratio $\left\langle K_1 \right\rangle / \left\langle K_2\right\rangle$. Values of $\mu_1$ are (left) $\mu_1 = 0.05$, (middle) $\mu_1 = 0.5$, and (right) $\mu_1 = 0.95$. Colors represent $\beta = 1, 2, 5$, and $10$ from blue to red. $\ell_1 = 1$.}
  \label{fig:aveK1-over-aveK2-vs-ell2}
\end{figure*}

\begin{figure}
  \centering
  \includegraphics[width=0.8\hsize]{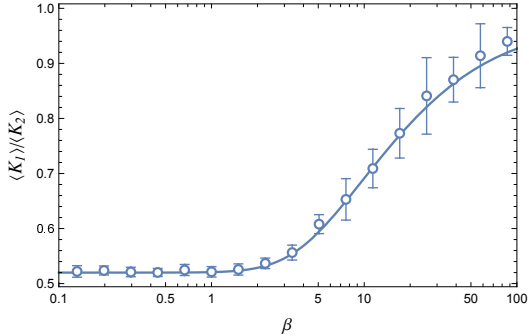}
  \caption{$\beta$-dependence of the ratio $\left\langle K_1\right\rangle/\left\langle K_2\right\rangle$. The solid line represents the values obtained from Eqs.(\ref{eq:ave-K1-result}) and (\ref{eq:ave-K2-result}), and the symbols represent values obtained by the Markov Chain Monte Carlo method.}
  \label{fig:aveK1-over-aveK2-vs-beta}
\end{figure}

Fig.~\ref{fig:Ki-ave-exact-and-mcmc} shows that $\beta \cdot \left \langle K_i \right \rangle $ appear to converge to certain finite values as $ \beta \rightarrow 0$. In fact, they are calculated as
\begin{align}
  \lim_{\beta\rightarrow 0}\beta\cdot  \left\langle K_1\right\rangle &
                                                                       = \frac{\mu_1}{2}\frac{K(\sqrt{\mu_2})}{E(\sqrt{\mu_2})} \ ,
\label{eq:aveK1-beta0}
\end{align}
and $\lim_{\beta\rightarrow 0}\beta\cdot\left\langle K_2\right\rangle$ is calculated from
Eq.(\ref{eq:ave-K2-result}). Here, $K(k)$ and $E(k)$ represent the complete elliptic integrals of
the 1st and 2nd kind~\cite{DLMF}, respectively. The calculation of Eq.(\ref{eq:aveK1-beta0}) is presented in Appendix \ref{sec:beta-0-limit}.
From this, we see that, the average kinetic energies do not depend on $\ell_i$ or $g$ at  the  high-temperature limit. This is similar to the approximate expression in Eq.(\ref{eq:ke-linear-tave-1}).

When $m_1 = m_2$, i.e., $\mu_1 = \mu_2 = 1/2$, we see that
\begin{align}
  \lim_{\beta\rightarrow 0}\beta\cdot  \left\langle K_1\right\rangle & = 0.3431\cdots \ ,\\
  \lim_{\beta\rightarrow 0}\beta\cdot  \left\langle K_2\right\rangle & = 0.6568\cdots \ .
\end{align}
These values agree well with the values obtained from the Markov Chain Monte Carlo method, as shown in Fig.~\ref{fig:Ki-ave-exact-and-mcmc}.

Now that we have obtained the exact expression for $\left\langle K_i\right\rangle$,
we can analyze their dependence on the parameters. %

In Fig.~\ref{fig:aveK1-over-aveK2-vs-ell2}, we show the $\ell_2$ dependence of the
ratio $\left\langle K_1 \right\rangle / \left\langle K_2\right\rangle$ for
$\mu_1 = 0.05$, $0.5$, and $0.95$. In every case, as
$\ell_2$ increases, that is, the length of the lower pendulum increases, 
the ratio $\left\langle K_1 \right\rangle / \left\langle K_2\right\rangle$
increases.

In Figure~\ref{fig:aveK1-over-aveK2-vs-beta}, we plot the $\beta$ dependence of the ratio $\left\langle K_1\right\rangle/\left\langle K_2\right\rangle$. We see that for $\beta\rightarrow 0$, the ratio converges to a positive value, and around $\beta \sim 1$, the ratio increases gradually. 

In these figures, we note two remarkable features:
\begin{list}{}{}
\item (i) %
At every temperature we have $\left\langle K_1\right\rangle < \left\langle K_2\right\rangle$.
That is, the particle at the end of the pendulum has a larger average kinetic energy than the other particle near the root.
\item (ii) The difference between the average kinetic energies is large for high temperatures and small for low temperatures, and a crossover temperature
is observed near $\beta = 1$.
\end{list}

The first feature is proved as follows. From Eq.(\ref{eq:trA1Ainv-dp}) and $\mu_1 + \mu_2 = 1$, we have 
\begin{equation}
  0 \le  \mbox{tr}\left( A^{(1)}A^{-1}\right)  \le 1 \le   \mbox{tr}\left( A^{(2)}A^{-1}\right)  \le 2 \ .
  \label{eq:trace-ineq}
\end{equation}
Eqs.(\ref{eq:trace-ineq}) and (\ref{eq:ave-Ki-linear-pi-trace}) indicate that
\begin{equation}
  0 \le  \left\langle K_1\right\rangle  \le \frac{1}{2\beta} \le \left\langle K_2\right\rangle  \le \frac{1}{\beta} \ 
  \label{eq:trace-ineq-ave}
\end{equation}
for any values of parameters $m_i$, $\ell_i$, and for any temperature.
Hence, the particle at the end always has a larger average kinetic energy than the other particle for a double pendulum.

\begin{figure}
  \centering
  \includegraphics[width=0.8\hsize]{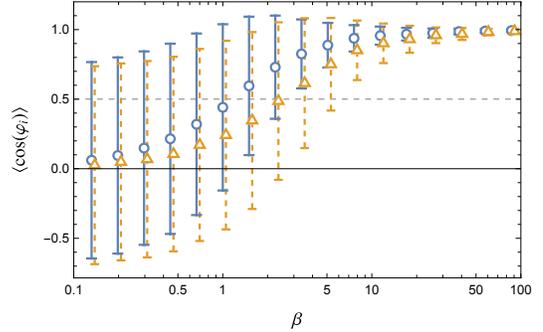}
  \caption{$\beta$-dependence of $\left\langle \cos(\varphi_i) \right\rangle$ obtained by the Markov Chain Monte Carlo method.  Open circles represent $i = 1$, and open triangles represent $i = 2$. The parameters were the same as those in Fig. \ref{fig:Ki-ave-exact-and-mcmc}. }
  \label{fig:ave-cos-varphi}
\end{figure}

Let us consider the latter feature. In Fig.\ref{fig:ave-cos-varphi}, we show $\left\langle \cos\varphi_i \right\rangle$ calculated by
the Markov Chain Monte Carlo method. We observe that for high temperatures $\beta \sim 0$,  $\left\langle \cos\varphi_i \right\rangle \sim 0$ for both $i = 1$ and $i = 2$.
Hence, angles take various values, and the pendulum shows the rotational motion.
For low temperature $\beta \gg 1$,
$\left\langle \cos\varphi_i \right\rangle \sim 1$
that is, $\varphi_i \sim 0$ for both $i = 1$ and $i = 2$; the pendulum exhibits a small-angle liberation.
Hence, the crossover temperature
observed in Fig.\ref{fig:Ki-ave-exact-and-mcmc} is related to crossover between the librational and rotational motions of the pendulum.

\begin{figure*}
  \centering
  \includegraphics[width=\hsize]{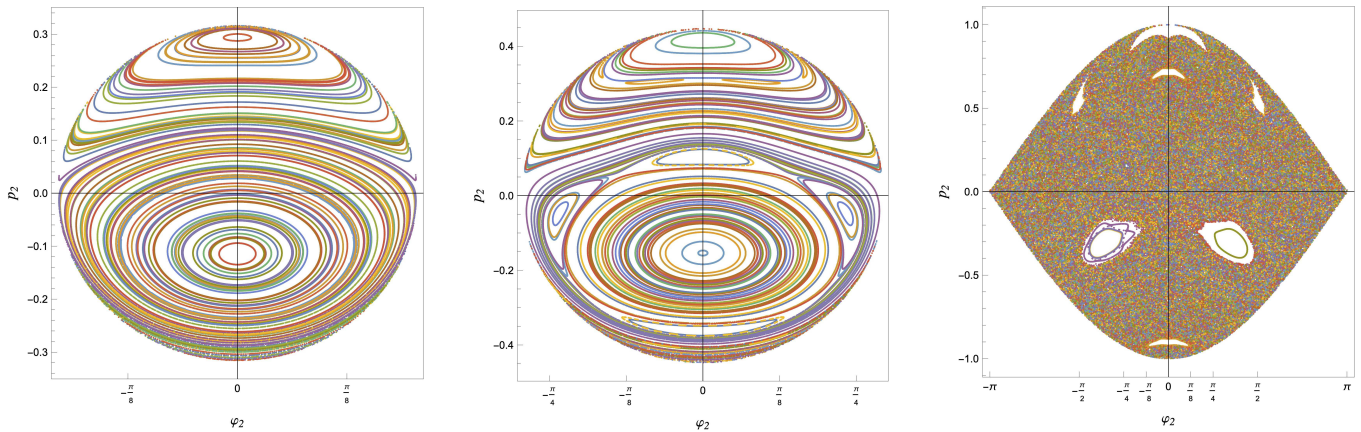}
  \caption{Poincar\'{e} surface of section $(\varphi_2,p_2)$ at $\varphi_1 = 0$, $p_1-p_2\cos\varphi_2 >0$. The definition of this section is explained in the Appendix \ref{sec:appendix-poincare-section-def}. 
$m_1 = m_2 = 1$, and $\ell_1 = \ell_2=1$. The values of the total energy are
(left) $E = 0.1$, \ (middle) $E = 0.2$, and \ (right) $E = 1.0$.  
}
  \label{fig:poincare-section}
\end{figure*}

The change in the motion was observed by examining the Poincar\'{e} surface of the section of the double pendulum. Fig.~\ref{fig:poincare-section} shows  the Poincar\'{e} surface of section $(\varphi_2,p_2)$  taken at
$\varphi_1=0$ and $p_1-p_2\cos\varphi_2 >0$ for several values of the total energy.
Definition of the section is explained in Appendix~\ref{sec:appendix-poincare-section-def}.
 As the total energy $E$ approaches $E\sim 1$, the 
chaotic region in the phase space rapidly expands and fills most of the energy surface. 
Hence, we interpret the crossover near $\beta~\sim 1$ found in previous figures as the change in motion from the limited librational to the strongly chaotic motion.

\section{Summary and discussions}
We showed that the averages of the linear kinetic energies  %
of particles in multiple pendulum take different values via numerical computation and analytical calculations.
Since the multiple pendulum has constraints, uniformity of average kinetic energy of each particle at thermal equilibrium is no longer guaranteed. On the contrary, another quantity what we call canonical kinetic energy is uniform, and the uniformity is guaranteed by the generalized principle of equipartition of energy. Moreover, the uniformity of the average canonical kinetic energy yields non-uniformity of the average kinetic energy of each particle.

Systems with constraints do not obey the (conventional) principle of the equipartition of energy, but a generalized one. Our discovery added a new example in which we can see how the system systematically deviates from (not generalized) equipartition~\cite{chain-letter-JSTAT-2009}.

The average linear kinetic energy of each particle was not the same 
because of the difference between the linear and canonical kinetic energies.
The difference 
comes from the fact that 
the kinetic energy
depends on coordinates; this dependence is attributed to the existence of 
constraints. In short, the constraints give rise to the nonuniform distribution of the average linear kinetic energies.
%
This scenario is similar to a chain system without a fixed root, i.e., a freely jointed chain.
Freely jointed chain is a simplified model of polymers and is  composed of particles connected by massless rigid links~\cite{fjc-Kuhn-1934,KUHN194811,Kramers-chain,Fixman3050,doi:10.1063/1.1681834,Mazers-chain-pre-1996,doi-edwards,doi-intro-polymer,strobl-polymer,SiegertWinklerReineker+1993+584+594}.
For a freely jointed chain,
the average kinetic energy of each particle
is large near both ends of the chain and small near the middle of the chain~\cite{chain-letter-JSTAT-2009}.

For a double pendulum, we successfully obtained the exact expressions 
for $\left\langle K_i\right\rangle$.
They include temperature, mass of two particles, lengths of two links,
and gravitational constant. Hence, the exact expression is useful to design a system that has predefined values of average kinetic energy. 
Exact expressions for  average kinetic energies are also obtained for
a three-particle two-dimensional freely jointed chain~\cite{konishi-yanagita-fjc-exact}, where non-uniformity of the average kinetic energies is also shown.

Further, we showed that the dependence of $\beta\cdot\left\langle K_i\right\rangle$ on the mass and length of links vanishes in the high-temperature limit. The reason why $\beta\cdot \left\langle K_i\right\rangle$ does not depend on $\ell_i$ or $g$ is interpreted as follows:  $\left\langle K_i\right\rangle$ depends on $\ell_i$ or $g$ only through coupling with $\beta$, as in Eq.(\ref{eq:alpha-g}) and (\ref{eq:alpha-ell}). 
Hence, in the limit $\beta\rightarrow 0$, the dependence on $\ell_i$ or $g$ vanishes.
This independence is also observed in the approximate expression in Eq.(\ref{eq:ke-linear-tave-1}).
Thus, the approximation adopted to obtain Eq.(\ref{eq:ke-linear-tave-1}) is similar to the high-temperature approximation.

In this paper we considered a multiple pendulum which have constraints, and the constraints
are essential for non-uniformity of average kinetic energies.
In real systems there are no exact constraints; constraints appearing in models
are often approximations of stiff springs and hard potentials.
Suppose we have a multiple spring-pendulum where the rigid links in the multiple pendulm
are replaced by springs.
If the potentials are sufficiently hard, there will be a  large gap between timescales of swinging motion of pendulum and vibrational motion of springs.
Then, according to Boltzmann-Jeans theory~\cite{boltzmann-1895,Jeans-1903,Jeans-1905,nakagawa-kaneko-pre-2001,benettin-galgani-giorgilli-pla-1987,benettin-jeans-2,benettin-jeans-ptp-94,chain-JSTAT-2010,chain-JSTAT-2016}, the energy exchange between 
swinging motion of pendulum and vibrational motion of springs take quite long time,
typically exponentiall long with respect to the spring constant.
Then we have a good chance to observe the system well approximated by rigid multiple pendulum, and average kinetic energies are non-uniform for quite long time.

Particles near the end of the pendulum have large 
average kinetic energy, which means the energy is localized to the particles
near the end of the pendulum.
There are other situations where the energy is localized to the 
end of the rope or chain; for example, the extremely high velocity of falling rope
and cracking whip~\cite{PhysRevLett.88.244301,falling-rope}. Whether these similar phenomena have the same dynamical origin is an interesting question.

\begin{acknowledgments}
  We would like to thank Mikito Toda and Yoshiyuki Y. Yamaguchi for their fruitful discussions.
  T. K. was supported by a Chubu University Grant (A).
T.Y. acknowledges the support of the Japan Society for the Promotion of Science (JSPS) KAKENHI Grants No. 18K03471 and 21K03411.
\end{acknowledgments}

\appendix
\section{Derivation of the Lagrangian and Hamiltonian of a multiple pendulum}\label{sec:lagrangian-general}
Here, we show a detailed derivation of the Lagrangian of 
a multiple pendulum~(Eq.(\ref{eq:lag-npend})).
To obtain a canonical momentum conjugate to the angle $\varphi_i$
\begin{equation}
  p_i \equiv \pdd{L}{\dot\varphi_i} \ \ ,
\end{equation}
we need to express $L$ (i.e., the kinetic energy $T$)
in terms of $\varphi_i$ and $\dot\varphi_i$.

First, we consider that the angles $\varphi$ and Cartesian coordinates $x,y$
are related as
\begin{align}
  x_i&= x_{i-1} + \ell_i \sin\varphi_i \ \
  = \sum_{j=1}^i \ell_j \sin\varphi_j \\
  y_i&= y_{i-1} - \ell_i \cos\varphi_i  \ \  
  = -\sum_{j=1}^i \ell_j \cos\varphi_j
\end{align}

Then, we have
\begin{align}
\dot x_i^2 + \dot  y_i^2
&= 
\left( 
\sum_{j=1}^i \ell_j \dot\varphi_j\cos\varphi_j
\right)^2
+
\left( 
 \sum_{j=1}^i \ell_j \dot\varphi_j\sin\varphi_j 
\right)^2 \nonumber \\
&= 
\sum_{j=1}^i
\sum_{k=1}^i
\ell_j\ell_k\dot\varphi_j\dot\varphi_k
\left(
\cos\varphi_j \cos\varphi_k + \sin\varphi_j\sin\varphi_k
\right) \nonumber
\\
&=
\sum_{j=1}^i
\sum_{k=1}^i
\ell_j\ell_k\dot\varphi_j\dot\varphi_k
\cos(\varphi_j -\varphi_k)
\end{align}
and the kinetic energy $K$ reads
\begin{equation}
  K =  \sum_{i=1}^N \frac{m_i}{2}
\sum_{j=1}^i
\sum_{k=1}^i
\ell_j\ell_k\dot\varphi_j\dot\varphi_k
\cos(\varphi_j -\varphi_k) \  .
\label{eq:ke-angle}
\end{equation}

Using an identity
\begin{equation}
  \sum_{i=1}^N \sum_{j=1}^i\sum_{k=1}^i 
=\sum_{j=1}^N \sum_{k=1}^N \left(\sum_{i=\max(j,k)}^N \right)\ ,
\end{equation}
we obtain
\begin{align}
  K &= \frac{1}{2}\sum_{j=1}^N \sum_{k=1}^N\left( \sum_{i=\max(j,k)}^N m_i\right)
\ell_j\ell_k\dot\varphi_j\dot\varphi_k
\cos(\varphi_j -\varphi_k) 
\\
&\equiv  
\frac{1}{2}\sum_{j=1}^N \sum_{k=1}^N A_{jk}(\varphi)
\dot\varphi_j\dot\varphi_k \ , 
\label{eq:K-total-matA}
\end{align}
where the $N\times N$ matrix $A(\varphi)$ is defined as
\begin{equation}
  A_{jk}(\varphi)=\left( \sum_{i=\max(j,k)}^N m_i\right)\ell_j\ell_k \cos(\varphi_j -\varphi_k) \ .
  \label{eq:mat-A-def}
\end{equation}

Using Eq.(\ref{eq:K-total-matA}), the Lagrangian of the multiple pendulum
is expressed in terms of the angles $\varphi$ and $\dot \varphi$ as 
\begin{align}
L
&=
\frac{1}{2}\sum_{j=1}^N \sum_{k=1}^N A_{jk}(\varphi)
\dot\varphi_j\dot\varphi_k 
+\sum_{i=1}^N m_i g \sum_{j=1}^i \ell_j\cos\varphi_j  \ , 
\label{eq:lag-angle-1}
\end{align}
where matrix $A$ is defined in Eq.(\ref{eq:mat-A-def}).

 The canonical momentum $p_n$ conjugate to the angle $\varphi_n$
($n = 1,2, \cdots, N$)
can be obtained using the expression of the Lagrangian (Eq.(\ref{eq:lag-angle-1})) as
\begin{align}
  p_n &\equiv \pdd{L}{\dot\varphi_n} = \sum_{k=1}^N A(\varphi)_{nk}\dot\varphi_k
\label{eq:p-and-phi}\\  
K &= \sum_{i,j=1}^N\frac{1}{2}A(\varphi)_{ij}\dot\varphi_i\dot\varphi_j 
\equiv \frac{1}{2}\dot\varphi^tA(\varphi)\,\dot\varphi
\label{eq:ke-mat-phi}\\
& = \sum_{i,j=1}^N\frac{1}{2}\left(A^{-1}(\varphi)\right)_{ij}p_i p_j \ .
\label{eq:ke-mat-p}
\end{align}

Then, the Hamiltonian of the multiple pendulum is obtained 
by the standard procedure as
\begin{align}
H &= \sum_{i=1}^n p_i\dot\varphi_i - L\nonumber \\
&=
\sum_{i,j=1}^N\frac{1}{2}\left(A^{-1}(\varphi)\right)_{ij}p_i p_j+ U(\varphi)
\end{align}

\section{``canonical kinetic energy''}\label{sec:app-canonical-ke}
 Our Hamiltonian has the form
 \begin{equation}
   H = \sum_{ij}\frac{1}{2}a_{ij}(q)p_i p_j + V(q)
   \label{eq:hamiltonian-with-off-diag}
 \end{equation}
where $q$ and $p$ are generalized coordinates and their conjugate momenta, respectively.

The generalized principle of the equipartition of energy is  expressed as 
\begin{equation}
  \left\langle \frac{1}{2}p_i \pdd{H}{p_i}\right\rangle = \frac{1}{2}k_B T  \ \ ( \text{no sum for $i$}) \ .
\end{equation}
Here, $k_B$ denotes the Boltzmann constant and $T$ represents the temperature.

$K_i^{(c)}$ is defined as
\begin{equation}
  K_i^{(c)} \equiv \frac{1}{2}p_i \pdd{H}{p_i}  \ \ ( \text{no sum for $i$})
\label{eq:ke-i-can}
\end{equation}

$K_i^{(c)}$  is not always the same as the traditional kinetic energy 
of the $i$'th degrees of freedom
\begin{equation}
  K_i \equiv \frac{1}{2}m_i \dot q_i^2 \ \  ( \text{no sum for $i$})
\label{eq:ke-i-lin}
\end{equation}
because of the off-diagonal terms in the quadratic form.
To clarify these distinction, 
we call $K_i^{(c)}$ the ``canonical kinetic energy'' 
and $K_i$ the ``linear kinetic energy'' in this paper.
\footnote{The term ``linear''
is used because the ``momentum'' $m\dot q$ is often called
as the ``linear momentum''.}

Below, we compute the ``canonical kinetic energy''
of the general multiple pendulum.

Our Hamiltonian  is of the form
\begin{equation}
  H(q,p) = K(q,p)+V(q)
\end{equation}
and we have for ``canonical kinetic energy,''
\begin{align}
  K_i^{(c)}
 &\equiv
\frac{1}{2}p_i \pdd{H}{p_i}
=  
\frac{1}{2}p_i \pdd{K}{p_i} \ \  (\text{no sum for $i$})
\\
&=
\frac{1}{2}p_i \pdd{}{p_i}
\sum_{j,k=1}^N\frac{1}{2}\left(A^{-1}\right)_{jk}p_j p_k \ \ 
(\text{from (\ref{eq:ke-mat-p})})\\
&=
\sum_{k=1}^N\frac{1}{2}p_i \left(A^{-1}\right)_{ik}p_k \ \ 
(\text{no sum for $i$}) \ \ .
\label{eq:ke-mat-p-i}
\end{align}

The sum of the ``canonical kinetic energy'' of the $i$'th 
angle 
$K_i^{(c)}$ is equal to 
the (total) kinetic energy $K$, which is given as
\begin{equation}
  K\equiv 
\sum_{i=1}^N\left(\frac{1}{2}p_i \sum_{k=1}^N\left(A^{-1}\right)_{ik}p_k 
\right)
=\sum_{i=1}^N K_i^{(c)} \ \ .
\end{equation}

Now, we express $K_i^{(c)}$ in terms of $\varphi$ and $\dot\varphi$,
and then in terms of the Cartesian coordinate $x, y$.
Using Eq.(\ref{eq:p-and-phi}), $K_i^{(c)}$ is expressed as
\begin{align}
K_i^{(c)}
&=
\sum_{j,k=1}^N\frac{1}{2}p_i \left(A^{-1}\right)_{ik}
\left(A_{kj}\dot\varphi_j
\right) \ \ (\text{no sum for $i$})\\
&= 
\frac{1}{2}p_i \dot\varphi_i 
\label{eq:ke-i-p-and-phi}\\
&=
\sum_{j=1}^N\frac{1}{2}A_{ij}\dot\varphi_j \dot\varphi_i \\
&=
\sum_{j=1}^N\frac{1}{2}\left(\sum_{n=\max(i,j)}^N m_n\right)
\ell_i \ell_j\cos(\varphi_i - \varphi_j) \dot\varphi_j \dot\varphi_i 
\label{eq:ke-can-i-phi-1}
\end{align}

This is the formula which represents ``canonical kinetic energy'' 
of the $i$'th degree of freedom in terms of $\varphi$ and $\dot\varphi$.

The Cartesian coordinates $x_i$ and $y_i$ express $K_i^{(c)}$.
The result is

\begin{align}
K_i^{(c)}
&=
\frac{1}{2}(\dot x_i -\dot x_{i-1})\left(\sum_{j=i}^N m_j \dot x_j\right)\nonumber\\
&+ 
\frac{1}{2}(\dot y_i -\dot y_{i-1})\left(\sum_{j=i}^N m_j \dot y_j \right) \ .
\label{eq:ke-can-i-xy-2}  
\end{align}

As expected, 
this expression for ``canonical kinetic energy'' $K_i^{(c)}$
is different from the ``linear'' kinetic energy
$K_i \equiv \frac{1}{2}m_i \left(\dot x_i^2 + \dot y_i^2\right) \ .
$
Although $K_i^{(c)}$ is defined from the momentum $p_i,$
which is canonically conjugate to the (local) angle $\varphi_i$,
$K_i^{(c)}$ is extended over all particles $j = 1, 2, \cdots, N$,
whereas $K_i$ is localized to a particular particle $i$.

\subsection{Example: Double pendulum}
Setting $N = 2$ in Eq.(\ref{eq:ke-can-i-xy-2}), 
we have the following expressions for the ``canonical kinetic energy''
of a double pendulum.
\begin{align}
  K_1^{(c)}
&=
\frac{1}{2}
\left\{
m_1 \left(\dot x_1^2 + \dot y_1^2\right)
+ m_2
\left(
\dot x_1 \dot x_2
+\dot y_1 \dot y_2
\right)
\right\}\\
K_2^{(c)}
&=
\frac{1}{2}m_2\left\{
(\dot x_2^2 + \dot y_2^2) - \left(\dot x_1 \dot x_2
+\dot y_1 \dot y_2
\right)
\right\}
\end{align}

We see that 
  \begin{align}
K_1^{(c)}&\ne K_1 \equiv \frac{1}{2}m_1\left(\dot x_1^2 + \dot y_1^2\right) \  , 
\\
K_2^{(c)}&\ne K_2 \equiv \frac{1}{2}m_2\left(\dot x_2^2 + \dot y_2^2\right) \ ,
\\
K_1^{(c)}+K_2^{(c)}&=K_1 + K_2 \ . 
  \end{align}

Therefore, at the thermal equilibrium, we have the average kinetic energy values of each particle of a double pendulum, which 
is not equal to $\frac{1}{2}k_B T$.
\begin{align}
\left\langle K_i\right\rangle &\ne \frac{1}{2} k_B T \ , i=1,2,\\
\left\langle K_1\right\rangle &\ne \left\langle K_2\right\rangle . 
\end{align}

\section{Calculation of $\left\langle K_i\right\rangle$ for multiple pendulum}\label{sec:Ki-ave-approx-stat}

Let us start with Eq.(\ref{eq:ave-Ki-linear-dotvarphi}).
By applying Eq.(\ref{eq:diagonal-approx-varphi}), we have
\begin{equation}
\left\langle K_i \right\rangle
=\frac{m_i}{2}
\sum_{j=1}^i
\ell_j^2 
\left<
\dot\varphi_j^2 
\right> \ .
\end{equation}

Now, let us evaluate $\left<\dot\varphi_j^2 \right>$ .
Note that
\begin{align}
  \dot\varphi_j &= \sum_{n=1}^N A^{-1}_{jn}p_n
                  = \frac{\partial}{\partial p_j}\frac{1}{2}\sum_{in}p_iA^{-1}_{in}p_n
                  = \frac{\partial}{\partial p_j} H
\end{align}

On the other hand,
\begin{equation}
  \pdd{}{p_j}e^{-\beta H} = -\beta\pdd{H}{p_j}e^{-\beta H} \ . 
\end{equation}

Therefore,
\begin{equation}
  \dot\varphi_j =\left(\frac{-1}{\beta}\right)  \frac{\partial }{\partial p_j}e^{-\beta H} 
\end{equation}
and 
\begin{align}
\int   \dot\varphi_j^2 e^{-\beta H} d\Gamma 
&=
 \left(\frac{-1}{\beta}\right) \int \left(\sum_n A^{-1}_{jn}p_n\right) \pdd{}{p_j}e^{-\beta H} d\Gamma \ .
\end{align}

Performing integration by parts with respect to $p_j$, we obtain
\begin{align}
&  \int \left(\sum_n A^{-1}_{jn}p_n\right) \pdd{}{p_j}e^{-\beta H} dp_j\\
&=
-
 \int  A^{-1}_{jj}(\varphi) \cdot e^{-\beta H} dp_j  \ , 
\end{align}
where the summation over $j$ is not considered.

Thus, we have
\begin{align}
\int   \dot\varphi_j^2 e^{-\beta H} d\Gamma 
&=
 \left(\frac{1}{\beta}\right) \int  A^{-1}_{jj}(\varphi)e^{-\beta H} d\Gamma \  , \\
\left< \dot\varphi_j^2 \right>
&=
\frac{1}{\beta}\left<  A^{-1}_{jj}(\varphi) \right> \ .
\end{align}

Let us evaluate $\left<  A^{-1}_{jj}(\varphi) \right>$.
From Eq.(\ref{eq:mat-A-def}),  we have
\begin{equation}
  A_{jj}(\varphi)\equiv\left(\sum_{i=j}^N m_i\right)\ell_j^2 \ .
\end{equation}

Let us adopt the approximation
\begin{align}
  A^{-1}_{jj}&\approx \frac{1}{A_{jj}} \\
  &= \frac{1}{\left(\sum_{i=j}^N m_i\right)\ell_j^2 } \ .
\end{align}
This approximation implies that we omit all nondiagonal elements of matrix $A$, which include phase factors such as $\cos(\varphi_{jk})$.

Then, we have
\begin{align}
\left< \dot\varphi_j^2 \right>
&=
\frac{1}{\beta}\left<  A^{-1}_{jj}(\varphi) \right>  
=\frac{1}{\beta}\left<\frac{1}{\left(\sum_{i=j}^N m_i\right)\ell_j^2 }\right>\\
&=
\frac{1}{\beta\left(\sum_{i=j}^N m_i\right)\ell_j^2 } \ .
\end{align}

Therefore, we obtain 
\begin{align}
\left< K_i \right>
&= \frac{m_i}{2}\left< \dot x_i^2 + \dot y_i^2 \right>
=  \frac{m_i}{2}\sum_{j=1}^i\left< \dot\varphi_j^2 \right>\ell_j^2 \\
&=
k_B T \cdot \frac{m_i}{2}\sum_{j=1}^i \frac{1}{\sum_{k=j}^N m_k} \  .
\end{align}

     \section{Calculation of $\lim_{\beta\rightarrow 0}\beta\cdot\left\langle K_1\right\rangle$ }\label{sec:beta-0-limit}
     Here, we present the calculation of  Eq.(\ref{eq:aveK1-beta0}), which represent the high-temperature limit   $\lim_{\beta\rightarrow 0}\beta\cdot\left\langle K_1\right\rangle $.
     From Eqs. (\ref{eq:ave-Ki-linear-pi-trace-varphi}), (\ref{eq:detA}), and (\ref{eq:trA1Ainv-dp}),  we have 
     \begin{equation}
       \beta    \left\langle K_1\right\rangle 
       =
         \frac{\mu_1}{2}
\frac{  \int_{(0,2\pi)^2}
  \frac{1}{\sqrt{1-\mu_2 C_{12}^2}} \, e^{-\beta U(\varphi)}d^2\varphi
}
         {\int_{(0,2\pi)^2} \sqrt{ 1-\mu_2 C_{12}^2}\, e^{-\beta U(\varphi)}d^2\varphi}
       \end{equation}
       for the double pendulum. Here, $C_{12} = \cos(\varphi_2-\varphi_1)$.
       Considering the limit $\beta\rightarrow 0$ on both sides, we have
       \begin{align}
         \lim_{\beta\rightarrow 0}\beta\cdot \left\langle K_1\right\rangle
&         = 
         \frac{\mu_1}{2}
                                                                             \frac{  \int_{(0,2\pi)^2} \frac{1}{\sqrt{1-\mu_2 C_{12}^2}}d^2\varphi} {\int_{(0,2\pi)^2} \sqrt{ 1-\mu_2 C_{12}^2}\, d^2\varphi}                                                                             \nonumber\\
         &=\frac{\mu_1}{2}
           \frac{K(\sqrt{\mu_2})}{E(\sqrt{\mu_2})} \ ,
           \label{eq:betaaveK1-beta0lim}
       \end{align}
       where $K(k)$ and $E(k)$ are the complete elliptic integrals of the 1st and 2nd kind, respectively.

\section{Poincar\'e surface of section}\label{sec:appendix-poincare-section-def}
Here, we explain the definition of the Poincar\'e surface of the section used in Fig. ~\ref{fig:poincare-section}.
A unique correspondence 
from a point $(\varphi_2,p_2)$  on the surface to a single point $(\varphi_1,\varphi_2,p_1,p_2)$,
in the phase space is necessary. We set two conditions $H(\varphi,p) =E $ and $\varphi_1 = 0$, which reduces the
four-dimensional phase space to a two-dimensional space $(\varphi_2,p_2)$. 
Let us specify a point $(\varphi_2,p_2)$ on the Poincar\'e surface.

For systems including a double pendulum, the energy is of the form
\begin{equation}
  \frac{1}{2}pA^{-1}p + U(\varphi_1,\varphi_2) = E  \ .
  \label{eq:2pend-energy}
\end{equation}
Let us express matrix $A$ as
\begin{equation}
  A =
  \begin{pmatrix}
    A_1 & A_2 \\ A_2 & A_3 
  \end{pmatrix}
  \ .
\end{equation}
By substituting this into Eq.(\ref{eq:2pend-energy}) and the straightforward calculation, we get 
\begin{equation}
     \left(A_3p_1 - A_2p_2\right)^2     = \det A\left\{      2A_3\left( E-U(\varphi_1,\varphi_2)\right) - p_2^2      \right\}
   \end{equation}
   Hence, point $(\varphi_2,p_2)$ corresponds to a unique point $(\varphi_1,\varphi_2,p_1,p_2)$ under $E = const$, and we must specify the sign of $A_3p_1 - A_2p_2$. Let us take a positive sign; then, the Poincar\'e surface of the section is defined as
   \begin{equation}
     \begin{cases}
       \varphi_1 &= 0 \ , \\
       A_3p_1 - A_2p_2 &> 0 \ .
       \label{eq:p-section-unique-1}
     \end{cases}
   \end{equation}

   In the case of a double pendulum, the last condition is equivalent to
   \begin{equation}
     \mu_2 \ell_2^2 p_1 - \mu_2 \ell_1\ell_2 p_2\cos\varphi_2  >0  \  .
     \label{eq:p-section-unique}
   \end{equation}
by substituting $\varphi_1 = 0$.
   For $\ell_1 = \ell_2$, this condition yields
   \begin{equation}
     p_1 -p_2 \cos\varphi_2  >0 \ , 
   \end{equation}
as shown in Fig.~\ref{fig:poincare-section}.

   Using
   \begin{equation}
     \dot\varphi = A^{-1} p,
   \end{equation}
   we can convert the condition in Eq.(\ref{eq:p-section-unique-1}) in terms of $ \dot \varphi$.
   Since 
   \begin{equation}
       A_3 p_1 - A_2 p_2 = \dot \varphi_1 \cdot \det A,
     \end{equation}
     and  the kinetic energy and matrix $A$  are positive definite,
     we can consider
     \begin{equation}
       \varphi_1 = 0 \ , \ \dot \varphi_1 > 0 
     \end{equation}
     as the surface of the section.

\bibliography{npend-paper}

\end{document}